\documentclass[prd,a4paper,showpacs,byrevtex,preprint,superscriptaddress]{revtex4}
\usepackage{graphicx}
\usepackage{dcolumn}
\usepackage{amsmath}
\usepackage{array}
\usepackage{bm}
\usepackage{amssymb}
\usepackage{amsfonts}
\usepackage{youngtab}
\begin{document}

\title{Spin contribution to light baryons in different large-$N$ limits}

\author{Fabien \surname{Buisseret}}
\email[E-mail: ]{fabien.buisseret@umons.ac.be}
\affiliation{Service de Physique Nucl\'{e}aire et Subnucl\'{e}aire,
Universit\'{e} de Mons--UMONS,
Acad\'{e}mie universitaire Wallonie-Bruxelles,
Place du Parc 20, 7000 Mons, Belgium}
\affiliation{Haute \'Ecole Louvain en Hainaut (HELHa), Chauss\'ee de Binche 159, 7000 Mons, Belgium}
\author{Nicolas \surname{Matagne}}
\thanks{F.R.S.-FNRS Postdoctoral Researcher}
\email[E-mail: ]{nicolas.matagne@umons.ac.be}
\author{Claude \surname{Semay}}
\thanks{F.R.S.-FNRS Senior Research Associate}
\email[E-mail: ]{claude.semay@umons.ac.be}
\affiliation{Service de Physique Nucl\'{e}aire et Subnucl\'{e}aire,
Universit\'{e} de Mons--UMONS,
Acad\'{e}mie universitaire Wallonie-Bruxelles,
Place du Parc 20, 7000 Mons, Belgium}

\date{\today}

\begin{abstract}
We investigate the spin contribution to light baryon ground states in three inequivalent large-$N$ limits: 't~Hooft, QCD antisymmetric and QCD symmetric. Our framework is a constituent quark model with a relativistic Hamiltonian containing a stringlike confinement and a one-gluon exchange term. Two spin-dependent potentials are considered and treated as perturbations: the color magnetic interaction stemming from the one-gluon exchange process and the chiral boson exchange interaction. We analytically prove that the spin contributions scale like $S(S+1)/n_q$, where $S$ is the total spin and $n_q$ is the number of quark, in agreement with diagrammatic methods. Both potentials yield also $S$-independent contributions which scale at most as O$(n_q)$.
\end{abstract}



\maketitle

\Yboxdim5pt

\section{Introduction}
\label{sec:intro}

The large-$N$ QCD approach is based on the replacement of the usual color group SU(3) by the group SU($N$) with a large arbitrary value $N$, allowing a perturbative expansion of the theory in $1/N$ \cite{hoof}. Relevant results can only be obtained if the real QCD ($N=3$) is not too different from the idealized world with large $N$. Surprisingly, it seems to be the case, taking into account the numerous successes obtained within this framework \cite{manohar}. Moreover, current lattice calculations also strongly support this idea (see e.g. the review \cite{teper}). 

In the original proposal by 't~Hooft \cite{hoof}, denoted here QCD$_\textrm{F}$, the quarks are in the fundamental representation of SU($N$) and the strong coupling constant $\alpha_S$ is such that the quantity $\alpha_0=\alpha_S\,N$ remains constant for any value of $N$. In this framework, a baryon is made of $N$ quarks in the totally antisymmetric color singlet and the number of flavors remains finite while $N\to\infty$. It has been shown by diagrammatic methods that the baryon mass thus scales as $N$ at the dominant order \cite{witten,luty93}. Later, Veneziano proposed a scheme in which the number of flavors grows like $N$ \cite{vene}. This new theory remains planar, but the internal quark loops are no longer suppressed as in the 't~Hooft limit.

Actually, the generalization of QCD to arbitrary numbers of colors is not unique, the main criterion being that the considered SU($N$) gauge theory has to be equivalent to QCD when $N=3$. For instance, a limit has been studied in which the quarks are in the two index antisymmetric representation of SU($N$), which is equivalent to the fundamental representation for $N=3$. Denoted here QCD$_\textrm{AS}$, that limit interestingly leads to a theory equivalent to ${\cal N}=1$ supersymmetric Yang-Mills when one light quark is present, as shown in \cite{qcdas}. In this framework, a baryon is made of $N(N-1)/2$ quarks in the totally antisymmetric color singlet \cite{bolo} and its mass scales as $N^2$ at the dominant order as shown in \cite{cohen} by diagrammatic methods, mostly for heavy quarks. In the same way, the quarks can also be considered in the two index symmetric representation of SU($N$) \cite{bolo}. Denoted here QCD$_\textrm{sym}$, this model is not equivalent to QCD$_\textrm{F}$ for $N=3$, but it is equivalent to some extent to QCD$_\textrm{AS}$ when $N\to\infty$ \cite{qcdas}. In this case, a baryon is made of $N(N+1)/2$ quarks in the totally antisymmetric color singlet and its mass is expected to scale as $N^2$ at the dominant order \cite{bolo}. Taking quarks in the two index symmetric representation is interesting since QCD-like theories with fermions in higher representations may be used in the so-called technicolor models \cite{ande11}. Other limits exit \cite{bolo2,corr,rytt06}, but as explained in Sec.~\ref{sec:barwfcolother}, they are not relevant for the model we use. So, they will not be considered in this work. 

The behavior with $N$ of several static properties of baryons (mass, size, contribution of strange quarks) have been predicted using nonrelativisctic potential models for heavy quarks systems \cite{witten}. Using diagrammatic methods, it was also suggested that the same results should hold for baryons made of light quarks. In the framework of a constituent quark model first suggested by Witten \cite{witten} (Hamiltonian with relativistic kinetic energy, stringlike confinement, and one-gluon-exchange term), we analytically proved that the static properties of light baryons scale as expected \cite{buis10}. These results has been obtained using the auxiliary field method (AFM) to obtain analytical upper and lower bounds of the $N$-body Hamiltonian considered \cite{afm,afmdual,afmnew,afmrev}. 

In this work, our purpose is to compute the $N$-behavior of the mass contribution for light baryons due to the spin $S$. We focus only on the ground states containing solely $u$ and $d$ quarks. The idea is to treat the spin-dependent interaction as a perturbation of the Hamiltonian used in our previous work \cite{buis10}. Three different limits are studied: QCD$_\textrm{F}$, QCD$_\textrm{AS}$, and QCD$_\textrm{Sym}$. Two spin-dependent potentials are considered: the color magnetic interaction stemming from one-gluon exchanges \cite{luch91} and the chiral boson exchange interaction \cite{gloz96}. In Sec.~\ref{sec:barwf}, we describe the baryon wavefunction for the different large-$N$ limits. The spin-independent Hamiltonian is presented in Sec.~\ref{sec:spinindH} with its approximate analytical solutions obtained in the framework of the AFM. The contributions of the two spin-dependent potentials are computed in Sec.~\ref{sec:spincont}. Some concluding remarks are given in Sec.~\ref{sec:conclu}. 

\section{Baryon wavefunction}
\label{sec:barwf}

The baryon wavefunction for $n_q$ quarks $|\phi\rangle$ is given by the product of a color part $|\phi_{C}\rangle$, a space part $|\phi_{X}\rangle$, and a flavor-spin part $|\phi_{FS}\rangle$
\begin{equation}
\label{totwf}
|\phi\rangle=|\phi_{C}\rangle\, |\phi_{X}\rangle\, |\phi_{FS}\rangle .
\end{equation}

\subsection{Color wavefunction}
\label{sec:barwfcol}

The color state of the baryon is a singlet one and it is taken completely antisymmetrical for the permutation of the $n_q$ quarks. This symmetry is imposed in order that the large-$N$ baryon is the most similar possible to a $N=3$ baryon (for QCD$_\textrm{AS}$ and QCD$_\textrm{Sym}$, several color singlets are possible \cite{bolo}). In this case, each pair of quarks is in the same antisymmetrical color configuration for the permutation of the two quarks. The color exchange operator $\bm \lambda^C_i \bm \lambda^C_j/4$ has the same value $F_{qq}$ for each pair, i.e.
\begin{equation}
\label{fqq}
F_{qq}= \frac{1}{2} \left( C_{qq} - 2\, C_q\right).
\end{equation}
In this relation, $C_q$ ($C_{qq}$) is the value of the quadratic Casimir operator for a quark (a pair of quarks). These values can be computed with standard formulas \cite{pere65,stancu}. The number $n_q$ depends on the color representation chosen for a single quark. We consider here 3 different limits, whose color parameters $n_q$, $C_q$, and $F_{qq}$ are gathered in Table~\ref{tab:colpar}. From this Table, it is clear that the QCD$_\textrm{F}$ and QCD$_\textrm{AS}$ schemes give identical numbers for $N=3$, while QCD$_\textrm{AS}$ and QCD$_\textrm{Sym}$ schemes give identical numbers for $N\to\infty$. As a pleasant feature, we have $F_{qq}=-2/3$ for the three cases when $N=3$.
\begin{table}[ht]
\centering
\caption{Color parameters for various large-$N$ limits.}
\begin{tabular}{cccc}
\hline\hline
 & QCD$_\textrm{F}$ & QCD$_\textrm{AS}$ & QCD$_\textrm{Sym}$ \\
\hline
$n_q$    & {\scriptsize $N$} & $\frac{N(N-1)}{2}$ & $\frac{N(N+1)}{2}$ \\
$C_q$    & $\frac{N^2-1}{2 N}$ & $\frac{(N-2)(N+1)}{N}$ & $\frac{(N-1)(N+2)}{N}$ \\
$F_{qq}$ & $-\frac{N+1}{2 N}$ & $-\frac{2}{N}$ & $-\frac{2}{N}$ \\
\hline\hline
\end{tabular}
\label{tab:colpar}
\end{table}

\subsubsection{QCD$_\textrm{F}$ limit}
\label{sec:barwfcolthooft}
\Yboxdim11pt

In the first limit proposed by 't~Hooft \cite{hoof}, the quark is in the fundamental representation of SU(N) $\yng(1)$ and it is necessary to gather $n_q=N$ quarks to form a color singlet. Each pair is in the antisymmetrical representation $\yng(1,1)$ and $F_{qq}=-(N+1)/(2 N)$. This limit has been extensively studied and has yielded numerous significant results \cite{manohar,teper,witten}.

\subsubsection{QCD$_\textrm{AS}$ limit}
\label{sec:barwfcolAS}

In this limit, the quark is in the representation $\yng(1,1)$ and an unique color singlet completely antisymmetrical exists with $n_q=\frac{N(N-1)}{2}$ \cite{bolo,cohen}. Each pair must be in the antisymmetrical representation for the exchange of two quarks. The only possibility is $\yng(2,1,1)$ whose dimension is $\frac{1}{8}(N-2)(N-1)N(N+1)$. One can check that this number is equal to $\frac{1}{2}n_q(n_q -1)$ and that $F_{qq}=-2/N$.

\subsubsection{QCD$_\textrm{Sym}$ limit}
\label{sec:barwfcolSym}

In this limit, the quark is in the representation $\yng(2)$ and an unique color singlet completely antisymmetrical exists with $n_q=\frac{N(N+1)}{2}$ \cite{bolo}. Each pair must be in the antisymmetrical representation for the exchange of two quarks. The only possibility is $\yng(3,1)$ whose dimension is $\frac{1}{8}(N-1)N(N+1)(N+2)$. One can check that this number is equal to $\frac{1}{2}n_q(n_q -1)$ and that $F_{qq}=-2/N$. Such a baryon may be called a ``technibaryon" in technicolor approaches \cite{ande11}. 

\subsubsection{Other limits}
\label{sec:barwfcolother}

In the Corrigan-Ramond limit \cite{corr}, baryons are three-quark states for any value of $N$: two quarks are in the fundamental representation and one is in the $(N-2)$-indice antisymmetric representation. We have shown in \cite{buis10}, that this limit cannot give relevant results for the main contribution to the baryon masses in the framework of a constituent model. So we choose not to investigate the Corrigan-Ramond limit for spin effects. Let us note that another limit which is somewhat in between the QCD$_\textrm{F}$ and Corrigan-Ramond ones has also been proposed \cite{rytt06}, but it is out of the scope of the present work since it requires a formalism in which quarks are Dirac spinors.

It has been also proposed a limit in which quarks are in the adjoint representation of SU($N$) (QCD$_\textrm{Adj}$) \cite{bolo2}. A stable skyrmion whose mass scales as O$(N^2)$ can exist within this framework, but the connection with the baryonic sector is not still clear. Such a limit presents also some unpleasant features: the adjoint representation has the dimension 3 for $N=2$, not for $N=3$; a color-singlet can already be obtained with only two quarks; no antisymmetrical color function with two adjoint quarks exists. Moreover, among all possible color functions for two adjoint quarks, none is such that $F_{qq}$ is negative and $\sim \textrm{O}(1/N)$. It is then impossible to obtain a baryon mass which scale as O$(N^2)$, as expected. So, we prefer not to investigate more this peculiar model.

\subsection{Space wavefunction}
\label{sec:barwfsp}

The auxiliary field method (AFM) can yield approximate solutions (eigenvalues and eigenvectors) for a system with an arbitrary number $n_q$ of identical particles \cite{afm,afmdual,afmnew,afmrev} (Most of the results presented in this section come from \cite{afm}, but we use here the more convenient notations developed in \cite{afmdual,afmnew,afmrev}). Within this method, the space part of an eigenstate is given by the product of $(n_q-1)$ oscillator states
\begin{equation}
\label{spacewf}
|\phi_{X}\rangle=\prod_{i=1}^{n_q-1}|n_i,l_i,\lambda_i,\bm x_i\rangle,
\end{equation}
where $n_i$ and $l_i$ are internal radial and orbital quantum numbers (the magnetic quantum numbers are omitted), and where $\bm x_i$ is a Jacobi coordinate and $\lambda_i$ is a scale parameter
\begin{equation}
\label{lambdai}
\lambda_i=\sqrt{\frac{i}{i+1}n_q\,Q}\frac{1}{r_0}.
\end{equation}
The global quantum number $Q$ is given by
\begin{equation}
\label{Q}
Q=K+\frac{3}{2}(n_q-1)\quad \textrm{with} \quad K=\sum_{i=1}^{n_q-1}(2\, n_i+l_i),
\end{equation}
where $K$ is the band number. The parameter $r_0$ is a kind of mean radius value for the system. It is the solution of a transcendental equation which depends on the kinematics, the interactions and the number of particles. For some well chosen structure of the Hamiltonian, an analytical form can be computed for $r_0$. Within this framework, each particle is characterized by a mean momentum $p_0$ given by
\begin{equation}
\label{p0}
p_0=\frac{Q}{r_0}.
\end{equation}
The state~(\ref{spacewf}) has neither a defined total angular momentum nor a good symmetry, but it is characterized by a parity $(-1)^K$. By combining such states with the same value of $K$, it is possible to build a physical state with good quantum numbers and good symmetry properties \cite{silv85}, but the task can be technically very complicated, even for $N=3$. Note that the ground state (GS) is given by
\begin{equation}
\label{GS}
|\phi_\text{X}(GS)\rangle=\prod_{i=1}^{n_q-1}|0,0,\lambda_i,\bm x_i\rangle,
\end{equation}
with $Q=\frac{3}{2}(n_q-1)$ and $\lambda_i=\sqrt{\frac{3 i}{i+1}\binom{n_q}{2}}\frac{1}{r_0}$, where the number of quark pairs $\frac{n_q(n_q-1)}{2}$ is denoted by the binomial coefficient $\binom{n_q}{2}$. Let us recall that
\begin{equation}
\label{binom}
\binom{m}{p}=\frac{m!}{p!(m-p)!}.
\end{equation}
The GS is a completely symmetrical positive parity S-state. 

\subsection{Flavor-spin wavefunction}
\label{sec:barwffs}

For the GS, the flavor-spin part $|\phi_{FS}\rangle$ of the wavefunction must be completely symmetrical. Flavor (F) and spin (S) are then associated with the same representation $[z]$ of the permutation group S$_{n_q}$ for $n_q$ particles, and we can write
\begin{equation}
\label{phiFS}
|\phi_{FS}\rangle = \frac{1}{\sqrt{d[\textrm{S}_{n_q}]_{[z]}}} \sum_{Y_z} |F [z] Y_z \rangle |S [z] Y_z \rangle,
\end{equation}
where the sum runs on the various standard tableaus indicated by $Y_z$. The numbers of these tableaus is the dimension $d[\textrm{S}_{n_q}]_{[z]}$ of the representation $[z]$ for the permutation group S$_{n_q}$. For baryons composed of $u$ and $d$ quarks only, this implies the equality of the spin ($S$) and the isospin ($T$), like for the baryons $N$ and $\Delta$ (in the following, we always use $S$ to present the results). The associated representation $[z]$ is \cite{luty93}
\Yboxdim11pt
\Yvcentermath1
\begin{equation}\label{young1}
[z]= 
\underbrace{\yng(1,1) \cdots \yng(1,1)}_{\frac{n_q-2S}{2}\ \textrm{columns}}
\raisebox{5.35pt}{$\hspace{-14.2pt}\overbrace{\yng(1)\cdots \yng(1)}^{2S\ \textrm{columns}}$}
\end{equation}
with $n_q$ boxes, where $2S$ and $n_q$ are odd numbers in order that the baryon is a fermion. $d[\textrm{S}_{n_q}]_{[z]}$ and the dimensions of this representation with respect to SU(2) [spin] and SU(3) [flavor] are given by:
\begin{align}
\label{SU23}
&d[\textrm{S}_{n_q}]_{[z]}=\frac{2S+1}{n_q +1} \binom{n_q +1}{(n_q-2S)/2}, \\
&d[SU(2)]_{[z]}= 2S+1, \\
&d[SU(3)]_{[z]}= \frac{(2S+1)(n_q-2S+2)(n_q+2S+4)}{8}.
\end{align}
These numbers can be computed by well known formulas \cite{stancu}. One can check that usual dimensions of the real QCD world are recovered with $n_q = 3$ for which $S=1/2$ or 3/2.

In the following, it will be necessary to determine the number of standard tableaus with the pair 1-2 in a symmetrical ($d[\textrm{S}_{n_q}]_{[z]}^\text{Sym}$) or antisymmetrical ($d[\textrm{S}_{n_q}]_{[z]}^\text{AS}$) state. This corresponds respectively to the following configurations: 
\begin{equation}\label{young2}
\young(12\cdot,\cdot\cdot\cdot)\cdots  \quad \textrm{and} \quad
\young(1\cdot\cdot,2\cdot\cdot)\cdots 
\end{equation}
\Yboxdim11pt
In the last case, the other numbers 3, 4, \ldots, $n_q$ are placed in the standard tableau in the same way as the numbers 1, 2, \ldots, $n_q-2$ can be placed in a tableau whose representation $[z']$ is the same as (\ref{young1}) but with the first column removed. So, $d[\textrm{S}_{n_q}]_{[z]}^\text{AS}$ is equal to $d[\textrm{S}_{n_q-2}]_{[z']}$. Obviously, $d[\textrm{S}_{n_q}]_{[z]}^\text{Sym}=d[\textrm{S}_{n_q}]_{[z]}-d[\textrm{S}_{n_q}]_{[z]}^\text{AS}$. We can then write more precisely 
\begin{equation}
\label{phiFS2}
|\phi_{FS}\rangle = \frac{1}{\sqrt{d[\textrm{S}_{n_q}]_{[z]}}} \left( \sum_{Y_z^\text{AS}} |F [z] Y_z^\text{AS} \rangle |S [z] Y_z^\text{AS} \rangle + \sum_{Y_z^\text{Sym}} |F [z] Y_z^\text{Sym} \rangle |S [z] Y_z^\text{Sym} \rangle\right),
\end{equation}
where the $d[\textrm{S}_{n_q}]_{[z]}^\text{AS}$ ($d[\textrm{S}_{n_q}]_{[z]}^\text{Sym}$) indexes $Y_z^\text{AS}$ ($Y_z^\text{Sym}$) correspond to configurations with a pair 1-2 antisymmetrical (symmetrical). Let us call $\beta$ the ratio
\begin{equation}
\label{alpha}
\beta=\frac{d[\textrm{S}_{n_q}]_{[z]}^\text{AS}}{d[\textrm{S}_{n_q}]_{[z]}}=\frac{n_q+2}{4(n_q-1)}-\frac{S(S+1)}{n_q(n_q-1)}.
\end{equation}
If $A_{12}$ is a spin-flavor operator acting on the pair 1-2 and $|\theta\rangle$ a $n_q$-body state with this pair such that $S_{12}=T_{12}=\theta$, then
\begin{equation}
\label{A12}
\langle \phi_{FS}| A_{12} |\phi_{FS}\rangle = \beta \langle 0| A_{12} |0 \rangle + (1-\beta) \langle 1| A_{12} |1 \rangle.
\end{equation}
The normality condition is recovered with $A_{12}=\openone$.
\Yboxdim5pt

\section{Spin-independent Hamiltonian}
\label{sec:spinindH}

The spin-independent Hamiltonian considered here was first proposed by Witten and it is used in \cite{buis10}, where justifications for its structure are detailed. It is written
\begin{equation}
\label{H0}
H_0=\sum_{i=1}^{n_q}\sqrt{\bm p^2_i}
+\frac{C_q}{C_{\yng(1)}}\sigma\sum_{i=1}^{n_q}\left|\bm r_i-\bm R\right|
+F_{qq}\, \frac{\alpha_0}{N}\sum_{i<j=1}^{n_q}\frac{1}{\left|\bm r_i-\bm r_j\right|}.
\end{equation}
The kinematic is semirelativistic with a vanishing mass for the quarks, since only $u$ and $d$ flavors are considered here. The confinement is insured by one-body linear terms and the one-gluon exchange interaction is taken into account (the one-boson exchange interaction will be presented in Sect.~\ref{sec:chiral}). The string tension $\sigma$ (0.15~GeV~$\lesssim \sigma \lesssim$~0.20~GeV) is $N$-independent up to corrections in $1/N^2$ \cite{make02}. $C_{\yng(1)}=\frac{N^2-1}{2 N}$ is the Casimir operator for the fundamental color representation. The strong coupling constant is given by $\alpha_0/N$ where $\alpha_0$ ($\alpha_0 \approx 1$) is the 't~Hooft coupling also independent of $N$. The color parameters $n_q$, $C_q$, and $F_{qq}$ appearing in this Hamiltonian are gathered in Table~\ref{tab:colpar} for the three large-$N$ limits considered here. From this Table, it is clear that the Hamiltonian is the same for the QCD$_\textrm{F}$ and QCD$_\textrm{AS}$ schemes for $N=3$, while it is the same for the QCD$_\textrm{AS}$ and QCD$_\textrm{Sym}$ schemes for $N\to\infty$.

The AFM equations can be solved analytically for Hamiltonian~(\ref{H0}), and $r_0$ is given by
\begin{equation}
\label{r0}
r_0=\sqrt{\frac{C_{\yng(1)}}{C_q\,\sigma} \left( n_q\, Q + \binom{n_q}{2}^{3/2}F_{qq}\, \frac{\alpha_0}{N} \right)}.
\end{equation}
So, within this approximation, the baryon masses are given by the following formula \cite{buis10}
\begin{equation}
\label{M0}
M_0=2\sqrt{\frac{C_q}{C_{\yng(1)}}\sigma\left( n_q\, Q + \binom{n_q}{2}^{3/2}F_{qq}\, \frac{\alpha_0}{N} \right)}.
\end{equation}
Formula~(\ref{M0}) is a generalization (to an arbitrary number of particles and to a non small value for $\alpha_0$) of relations obtained in \cite{barn3}. It is an upper bound, but it was shown in \cite{buis10} that the lower bound for the GS has the same behavior in $N$. So we are confident about the results for $N\to \infty$ obtained by this kind of formula. 

\begin{figure}[ht]
\begin{center}
\includegraphics*[width=10cm]{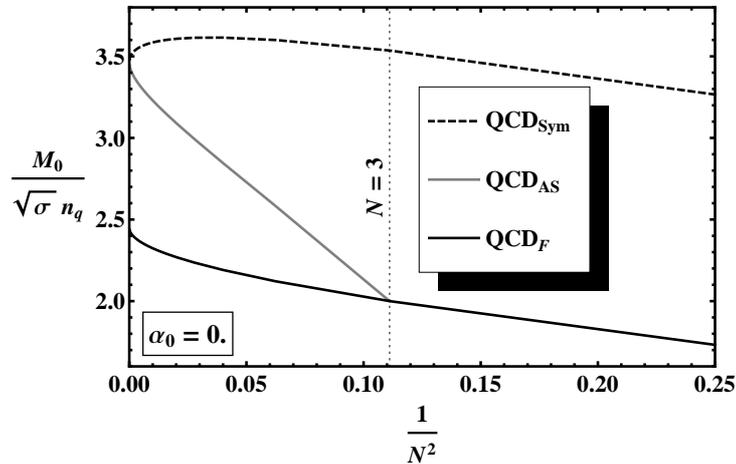}
\caption{Baryon GS masses per quark in $\sqrt{\sigma}$ unit for three large-$N$ limits, with $\alpha_0=0$. In order to guide the eyes, $N$ is considered as a continuous variable.}
\label{fig:baryon}
\end{center}
\end{figure}

More precisely, the dominant contributions for the GS within the various limits $N\to \infty$ are:
\begin{align}
\label{MNinfty1}
&M_0^{\textrm{QCD$_\textrm{F}$}}=N\sqrt{\sigma\left( 6-\frac{\alpha_0}{\sqrt{2}} \right)}, \\
\label{MNinfty2}
&M_0^{\textrm{QCD$_\textrm{AS}$}}=M_0^{\textrm{QCD$_\textrm{Sym}$}}=\frac{N^2}{2}\sqrt{\sigma\left( 12-2\sqrt{2}\,\alpha_0 \right)}.
\end{align}
As expected and already shown in \cite{buis10}, $M_0$ always scales as $n_q$, although this point is non trivial in the light quark case. The reduced baryon GS masses per quark for finite $N$ values are given in Fig.~\ref{fig:baryon} for $\alpha_0=0$. For finite value of $\alpha_0$, the behavior is similar but the masses are lowered. It has to be remarked that QCD$_\textrm{AS}$, QCD$_\textrm{Sym}$, and QCD$_\textrm{Adj}$ theories can be shown to be equivalent at large $N$ \cite{qcdas}. This so-called orientifold equivalence a priori holds for charge-conjugation invariant sectors of the theory. While the meson masses are logically found to be compatible in the QCD$_\textrm{AS}$, QCD$_\textrm{Sym}$, and QCD$_\textrm{Adj}$ limits \cite{luci10}, the present result was not expected a priori since the baryonic sector is not charge-conjugation invariant \cite{unsa06}. 

Only the dominant contribution to the baryon masses is given by (\ref{MNinfty1}) and (\ref{MNinfty2}). Several corrections are expected from different sources. Parameters $\sigma$ and $\alpha_0$ are not supposed to be real constants with respect to $N$. All interactions, depending on spin or spin-orbit for instance, are certainly not included in the Hamiltonian~(\ref{H0}). Other quantum mechanisms, such as self-interaction, are not easily simulated by potential models. Nevertheless, none of these effects is expected to modify strongly the dominant scaling expressed by (\ref{MNinfty1}) and (\ref{MNinfty2}) \cite{witten}. Moreover, the presence of $n_s$ ($n_s \ll n_q$) strange quarks with a small mass ($m_s < \sqrt{\sigma}$) can be treated as a perturbation and does not modify the dominant mass scale of the baryon \cite{buis10}. 

\section{Spin contributions}
\label{sec:spincont}

The spin contributions to baryons considered here are two-body forces $W_{ij}$ and will be treated as perturbations. As the GS is completely symmetrized, we can write (see Appendix~\ref{app:meanval})
\begin{equation}
\label{Wspin}
\langle W\rangle=\sum_{i<j=1}^{n_q}\langle W_{ij}\rangle=\binom{n_q}{2} \langle W_{12}\rangle.
\end{equation}
The pair 1-2 is chosen because $\bm r_1 - \bm r_2=\bm r_{12}=\bm x_1$ \cite{afm}, and the use of only one Jacobi coordinate simplifies greatly the calculations. In the real world, the mean mass of the $\Delta$ and the nucleon is around 1~GeV, while $m_\Delta - m_N\approx 0.3$~GeV. So we can ask if the perturbation treatment is really pertinent. Moreover, these spin interactions vary as $1/m^2$ where $m$ is the quark mass: 
with the vanishing mass used here, the interaction should blow up. How to cure these potential flaws of the model?

First, in the large-$N$ limits, the spin-independent mass scales as $n_q$, while the spin correction is expected to scale as $1/n_q$. So, the spin contribution must become less and less large as $N$ increases, justifying a priori a perturbative treatment. Second, the $1/m^2$ dependence comes from a genuine nonrelativistic limit of the exchange diagrams. A better approximation should be to replace the term $1/m$ by the operator $1/\sqrt{\bm p^2+m^2}$ which is always finite, even for vanishing quark masses \cite{luch91}. As this operator is very complicated to use, it will be replaced in our work by $1/\mu_0$, where $\mu_0$ is the auxiliary field associated with the kinetic part of the Hamiltonian. This quantity is such that $\mu_0^2=\langle \bm p^2+m^2\rangle$ for the state considered. It can be shown that $\mu_0=\sqrt{p_0^2+m^2}$ with $p_0$ given by (\ref{p0}) \cite{afm}.

In mass formulas issued from large-$N$ works, the spin contribution is proportional to $\bm S^2$, with $\bm S=\sum_{i=1}^{n_q} \bm s_i$, while the sum on pairwise operators $\bm s_i\cdot\bm s_j$ appears in potential models. The link between these two approaches is given by
\begin{equation}
\label{S2}
\sum_{i<j=1}^{n_q} \bm s_i\cdot\bm s_j = \frac{1}{2} \left( \bm S^2 - \sum_{i=1}^{n_q} \bm s_i^2\right).
\end{equation}

\subsection{One-gluon exchange}
\label{sec:OGE}

The spin-spin interaction coming from the one-gluon exchange (OGE) interaction is well known and its form can be found in many textbooks (see for instance \cite{gloz96,luch91}). As explained above, we replace the $1/m^2$ term by $1/\mu_0^2$. With constant factors replaced by a generic constant $A=8\pi/3$ and the strong coupling constant $\alpha_S$ replaced by $\alpha_0/N$, the form retained is  
\begin{equation}
\label{OGE}
W_{12}^\textrm{OGE}=-\frac{A}{\mu_0^2}\, \frac{\alpha_0}{N}\, \delta^3(\bm r_{12})\, F_{12}\, \bm s_1\cdot\bm s_2.
\end{equation}
As this potential is treated as a perturbation, the delta distribution gives a finite result. Note that to reproduce correctly the spin splitting with the OGE mechanism, it seems necessary to take into account the running character of the coupling constant and subtle relativistic corrections \cite{godf85}. Such detailed features are not relevant for our purpose.

The various factors of $\langle W_{12}^\textrm{OGE}\rangle$ can now be computed. In our case, $\mu_0=p_0=Q/r_0$ with $r_0$ given by (\ref{r0}). All the color factors are equal, so $F_{12}=F_{qq}$. For the GS, the mean value of the Dirac distribution is given by 
\begin{equation}
\label{delta12}
\langle\phi_\text{X}(GS)|\delta^3(\bm r_{12})|\phi_\text{X}(GS)\rangle=\langle 0,0,\lambda_1,\bm x_1|\delta^3(\bm x_1) |0,0,\lambda_1,\bm x_1\rangle
=\frac{\lambda_1^3}{\pi^{3/2}},
\end{equation}
with $\lambda_1$ given by (\ref{lambdai}). The last result is obtained using the properties of the harmonic oscillator. Using (\ref{Wspin}) and (\ref{S2}), it is easy to show that
\begin{equation}
\label{s12}
\langle \bm s_1\cdot\bm s_2\rangle=\frac{4 S(S+1)-3 n_q}{8 \binom{n_q}{2}}.
\end{equation}
This relation can also be obtained from (\ref{alpha}) and (\ref{A12}). Though we only consider states with $S=T$, formula (\ref{s12}) is due to the spin number $S$ alone. Putting all the results together, we find
\begin{equation}
\label{WOGE}
\langle W^\textrm{OGE} \rangle=-\frac{\alpha_0 A}{8 \sqrt{6}\pi^{3/2}} \frac{F_{12}}{N} \frac{n_q}{n_q-1} \sqrt{\frac{\frac{C_q}{C_{\yng(1)}}\sigma}{3 + \binom{n_q}{2}^{1/2}F_{12}\, \frac{\alpha_0}{N}}} \left[4 S(S+1)-3 n_q\right].
\end{equation}
Using the values in Table~\ref{tab:colpar}, the spin contributions for the various limits can be computed. In particular the dominant contributions for $N\to\infty$ are given by:
\begin{align}
\label{spininfty1}
&\langle W^\textrm{OGE}_{\textrm{F}} \rangle= \frac{\alpha_0 A}{2 \pi^{3/2}}\sqrt{\frac{\sigma}{6(12 - \sqrt{2}\,\alpha_0)}} \left[ \frac{S(S+1)}{N} - \frac{3}{4}\right] + \textrm{O}\left( \frac{1}{N} \right),\\
\label{spininfty2}
&\langle W^\textrm{OGE}_{\textrm{AS}} \rangle =\langle W^\textrm{OGE}_{\textrm{Sym}} \rangle=\frac{\alpha_0 A}{\pi^{3/2}}\sqrt{\frac{\sigma}{6(6 - \sqrt{2}\,\alpha_0)}} \left[ \frac{S(S+1)}{N^2/2} - \frac{3}{4}\right] + \textrm{O}\left( \frac{1}{N} \right).
\end{align}
Only the dominant $S$-dependent and $S$-independent contributions are indicated. We consider that higher order corrections are not fully relevant since some parameters are given at their dominant order only ($\alpha_0$ and $\sigma$). For each limit, we see that $\langle W^\textrm{OGE} \rangle \propto S(S+1)/n_q$ for $N\to\infty$, as expected. But, it appears other terms which are $S$-independent. These particulars contributions are unavoidable in the framework of a potential model. They are not really disturbing to fit the parameters of the mass formulas since they can be absorbed in the various terms of the large-$N$ expansion. Such a large-$N$ behavior was already shown for nonrelativistic constituent quark models \cite{hida11}. 

The sign of $\langle W^\textrm{OGE} \rangle$ is controlled by the factor $\left[4 S(S+1)-3 n_q\right]$ since $F_{12}$ is always negative. The contribution due to the spin always increases with $S$: for $S=1/2$, it is negative; for $S=3/2$, it is strictly positive only for $n_q=3$, and for the maximum value $S=n_q/2$, it is positive. The dominant contribution of the spin is of order O(1) for all large-$N$ limits. As, the baryon mass is at least of order O$(N)$, a perturbative calculation is justified when $N\to\infty$. 

\subsection{Chiral boson exchange}
\label{sec:chiral}

The chiral boson exchange (CBE) mechanism has been proposed as an alternative to the one-gluon exchange process \cite{gloz96,gloz98}. Based on the approximate chiral symmetry of QCD, it can yield very good baryon spectra. In particular, it is possible to solve the hierarchy problem between the positive and negative parity partners for excited nucleons. As our purpose is only to check the $N$-dependence of this interaction, we will consider the simplest representation of the most important component of the potential that is mediated by the octet of pseudoscalar bosons. In the SU(3)$_F$ invariant limit, we can write
\begin{equation}
\label{CBE}
W_{12}^\textrm{CBE}=\frac{B}{\mu_0^2}\, g^2\, V^\textrm{CBE}(\bm r_{12}) \, \bm s_1\cdot\bm s_2\, \frac{\bm \lambda^F_1\cdot\bm\lambda^F_2}{4},
\end{equation}
with the constant $B=1/(3\pi)$ and the usual $1/m^2$ term replaced by $1/\mu_0^2$. The operator $\bm \lambda^F_1\cdot\bm\lambda^F_2/4$ is acting in the flavor space, and the radial part of this interaction is given by 
\begin{equation}
\label{CBEVr}
V^\textrm{CBE}(\bm x)=\Lambda^2 \frac{e^{-\Lambda x}}{x}-4\pi \delta^3(\bm x).
\end{equation}
Again, as this potential is treated as a perturbation, the delta distribution gives a finite result. The parameter $g$ is a coupling constant whose $N$-dependence is discussed below. The quantity $\Lambda$ is the common mass for the pseudoscalar bosons in this simplified model. So, we can expect that $m_\pi \le \Lambda \le m_\eta$. As a meson mass is, at the dominant order, constant within large-$N$ models, this parameter is assumed to be independent of $N$. Note that in this framework, no gluon exchange has to be considered. So, we set $\alpha_0=0$ in (\ref{H0}) for this model.

The coupling constant $g$ is given by the following formula \cite{gloz96}
\begin{equation}
\label{gCBE}
g= m_u \frac{g_A}{f_\pi},
\end{equation}
where $g_A$ is the quark vector axial coupling constant, $f_\pi$ is the pion decay constant, and $m_u$ the effective mass of the quark $u$. In the context of our semirelativistic Hamiltonian (\ref{H0}), we identify $m_u$ with $\mu_0$. So $m_u^2$ in $g^2$ simplifies with the $1/\mu_0^2$ quantity in (\ref{CBE}). The problem is to find the $N$-dependence of the remaining factor in $g$. In Appendix~\ref{app:ga}, we give some arguments supporting the fact that $f_\pi^2 \sim \textrm{O}(n_q)$ in agreement with \cite{bron93,sann07}, and that $g_A \sim \textrm{O}(1)$ in agreement with \cite{hida11,wein90,bron93}. So, we propose to take $g_A$ as a pure constant, and to set  
\begin{equation}
\label{fpi}
f_\pi (N) = \sqrt{\frac{n_q}{3}} f_\pi(3),
\end{equation}
where $f_\pi(3)=131$~MeV. This normalization insures a correct physical value when $N=3$ for both QCD$_{\textrm{AS}}$ and QCD$_{\textrm{F}}$ since $n_q=3$ in these cases. This seems compulsory for QCD$_{\textrm{AS}}$ which is equivalent to QCD$_{\textrm{F}}$ for $N=3$. On the contrary, there is no strong argument to impose such an equivalence between QCD$_{\textrm{Sym}}$, for which $n_q=6$, and QCD$_{\textrm{F}}$ (see Fig.~\ref{fig:baryon}). It seems preferable to preserve the equivalence between QCD$_{\textrm{AS}}$ and QCD$_{\textrm{Sym}}$ when $N\to\infty$.

For the GS, the mean value of the radial potential is given by ($\bm r_{12} = \bm x_1$)
\begin{eqnarray}
\label{W12}
\lefteqn{\langle \phi_\text{X}(GS)|V^\textrm{CBE}(\bm x_1)|\phi_\text{X}(GS)\rangle =
W(\Lambda,\lambda_1)} \nonumber \\ &=& \frac{2\lambda_1(\Lambda^2-2\lambda_1^2)}{\sqrt{\pi}}-\Lambda^3 \exp\left( \frac{\Lambda^2}{4\lambda_1^2}\right) \textrm{erfc}\left( \frac{\Lambda}{2\lambda_1}\right).
\end{eqnarray}
with $\lambda_1$ given by (\ref{lambdai}). The last result is obtained using the properties of the harmonic oscillator. The sign of such a quantity is not obvious. It can be checked that for physical (positive) values of the arguments, $W(\Lambda,\lambda_1)$ is always negative, even if its magnitude can largely varies. 
Using (\ref{alpha}) and (\ref{A12}), one obtains 
\begin{equation}
\label{f12s12}
\left\langle \frac{\bm \lambda^F_1\cdot\bm\lambda^F_2}{4} \bm s_1\cdot\bm s_2\right\rangle=\frac{3 n_q(3 n_q+2) - 20 S(S+1)}{96 \binom{n_q}{2}}.
\end{equation}
This formula is in agreement with the results from \cite{gloz96} for $n_q=3$ and both $S=1/2$ and 3/2. The spin-flavor operator of the CBE interaction acts in such way that it is not possible to disentangle the contributions from spin and isospin. Remember that we consider here only states with $S=T$.
Putting all the results together, we find
\begin{equation}
\label{WCBE}
\langle W^\textrm{CBE} \rangle=\frac{g_A^2\,B}{32\, f_\pi^2(3)}\frac{1}{n_q} W\left(\Lambda,\sqrt{\frac{C_q \, \sigma}{2\, C_{\yng(1)}}}\right) [3 n_q(3 n_q+2) - 20 S(S+1)].
\end{equation}
Using the values in Table~\ref{tab:colpar}, the spin contributions for the various limits can be computed. In particular the dominant contributions for $N\to\infty$ are given by:
\begin{align}
\label{spininfty1b}
&\langle W^\textrm{CBE}_{\textrm{F}} \rangle= \frac{5 g_A^2 B}{8 f_\pi^2(3)} \left|W(\Lambda,\sqrt{\sigma/2})\right|\left[ \frac{S(S+1)}{N} - \frac{9}{20} N\right] + \textrm{O}(1),\\
\label{spininfty2b}
&\langle W^\textrm{CBE}_{\textrm{AS}} \rangle =\langle W^\textrm{CBE}_{\textrm{Sym}} \rangle=\frac{5 g_A^2 B}{8 f_\pi^2(3)} \left|W(\Lambda,\sqrt{\sigma})\right| \left[ \frac{S(S+1)}{N^2/2} - \frac{9}{20}\frac{N^2}{2}\right] + \textrm{O}(N),
\end{align}
where it is taken into account that $W(\Lambda,\lambda_1)$ is always a negative number. Only the dominant $S$-dependent and $S$-independent contributions are indicated. We consider that higher order corrections are not fully relevant since some parameters are given at their dominant order only ($g_A$, $f_\pi(N)$, $\Lambda$, and $\sigma$). For each limit, we see again that $\langle W^\textrm{CBE} \rangle \propto S(S+1)/n_q$ for $N\to\infty$ and that terms appears which are $S$-independent. Again, they are not really disturbing to fit the parameters of the mass formulas since they can be absorbed in the various terms of the large-$N$ expansion.  

The sign of $\langle W^\textrm{CBE} \rangle$ is controlled by the factor $\left[20 S(S+1)-3 n_q(3 n_q+2)\right]$ since $W(\Lambda,\lambda_1)$ is always negative. The contribution due to the spin always increases with $S$ but it is negative from $S=1/2$ to $S=n_q/2$. The dominant contribution of the CBE interaction and the baryon mass are both of order O$(n_q)$. Following the above results, we have:
\begin{align}
\label{CBEpert1}
&M_0 \sim n_q \sqrt{6\, \sigma},\\
\label{CBEpert2}
&\langle W^\textrm{CBE} \rangle \sim n_q \frac{3\, g_A^2 \left|W(\Lambda,\sqrt{\sigma})\right|}{32\, \pi f_\pi^2(3)}.
\end{align}
For reasonable values of the parameters, the ratio $\langle W^\textrm{CBE} \rangle/M_0$ is around 30\%, as for the experimental case. Contrary to the situation for the OGE potential, the large-$N$ limit does not reduce the magnitude of the CBE interaction with respect to the dominant Hamiltonian (\ref{H0}). So, a perturbative calculation seems less justified for this spin-dependent potential. 

\subsection{Summary}
\label{sec:summary}

The balance sheet of the various effects leading to the universal behavior $\bm S^2/n_q$, whatever the spin-spin interaction, is interesting to make. The contribution of the integral over the space part of the spin-spin potential is of order $\textrm{O}(1)$. This is expected since a baryon with Hamiltonian (\ref{H0}) has a quasi constant size with $N$ \cite{buis10}. The contribution of the quark effective mass $\mu_0$ is also of order $\textrm{O}(1)$. This is also not a surprise since the mass of the baryon, which scale as $n_q$, is always proportional to $n_q\, \mu_0$. The (flavor-)spin operator for the pair 1-2 is always proportional to $1/\binom{n_q}{2}$ and this factor is compensated by the factor $\binom{n_q}{2}$ coming from the sum over all the pairs. Finally, the $N$-dependence comes essentially from the remaining factor: the strong coupling constant times the color operator, $F_{12}\,\alpha_0/N$, for the OGE; the coupling constant $g^2$ for the CBE. Despite their very different origins, both quantities scales as $1/n_q$ for each large-$N$ limit considered here.

\section{Concluding remarks}
\label{sec:conclu}

In this paper, we have computed the $N$-behavior of the mass contribution for the light baryons due to the spin $S$, in the framework of a constituent quark model. The spin-independent Hamiltonian considered here was first proposed by Witten \cite{witten} and used in \cite{buis10} where it is fully described. The kinematic is semirelativistic with a vanishing mass for the quarks, since only $u$ and $d$ flavors are considered. The confinement is simulated by a sum of one-body linear terms and the one-gluon exchange interaction is taken into account. The idea is to treat the spin-dependent interaction as a perturbation of this Hamiltonian. All computation are approximate but analytical thanks to the use of the auxiliary field method \cite{afm,afmdual,afmnew,afmrev}. Two spin-dependent potentials are considered: the color magnetic interaction stemming from one-gluon exchanges \cite{luch91} and the chiral boson exchange interaction \cite{gloz96}. 

Three different large-$N$ limits are studied. In the first limit proposed by 't~Hooft \cite{hoof}, a quark is in the fundamental representation (QCD$_\textrm{F}$) of SU(N) and it is necessary to gather $n_q=N$ quarks to form a color singlet. The second considered here is the antisymmetric one in which quark are in the two index antisymmetric representation (QCD$_\textrm{AS}$) of SU($N$) and form an unique color singlet completely antisymmetrical with $n_q=N(N-1)/2$ \cite{bolo,cohen}. In the last one studied here, a quark is in the two index symmetric representation (QCD$_\textrm{Sym}$) of SU($N$) and an unique color singlet completely antisymmetrical exists with $n_q=N(N+1)/2$ \cite{bolo}.

Even if approximate analytical results are obtained, we have shown in \cite{buis10} that the $N$-behavior of the solutions are correct when $N\to\infty$. Within our model, the results are the same for the QCD$_\textrm{F}$ and QCD$_\textrm{AS}$ schemes for $N=3$, while they are the same for the QCD$_\textrm{AS}$ and QCD$_\textrm{Sym}$ schemes for $N\to\infty$. This is in agreement with the results in \cite{cher2}, where it has been shown that predictions for baryon mass relations obtained with QCD$_\textrm{F}$ and QCD$_\textrm{AS}$ limits are both in agreement with experimental data. 

We focus only on the ground states containing solely $u$ and $d$ quarks, and thus global spin $S$ and global isospin $T$ of the baryons are the same for symmetry reasons. As we consider only $S=T$ states, it is not possible to disentangle the contributions coming from spin and isospin. However, it could be interesting to study in future works other states with $S\ne T$, since isospin-dependent operators could play an important role for the masses of some multiplets \cite{mata08}. 

The main result of this work is that the $S$-dependent mass term for light baryons is proportional to $S(S+1)/n_q$ when $N\to \infty$, as already shown from diagrammatic methods, mostly valid for heavy quarks. It is obtained for both the one-gluon exchange mechanism and the chiral boson exchange potential. Despite their different origins, both contributions are characterized by a strength varying as $1/n_q$ for each large-$N$ limit considered here. These interactions yield also $S$-independent contributions which behave very differently. For the one-gluon exchange mechanism, the corresponding contribution scales as O(1). So a perturbative treatment is fully justified when $N\to\infty$, since the baryon mass scale as O($n_q$). For the chiral boson exchange potential, the corresponding contribution scales as O($n_q$). In this case, the limit $N\to\infty$ does not influence the quality of the perturbative treatment. From our point of view, it is not possible to prefer one interaction with respect to the other on the basis of our results. For instance, the $S$-independent contributions can be absorbed in various terms of baryon mass formulas. We prefer to focus on the fact that both potentials gives the expected term proportional to $S(S+1)/n_q$. We think that this work and the previous one \cite{buis10} validate our approach to study baryons in various large-$N$ limits. Since approximate analytical baryon eigenfunctions are available with our method, a lot of observables can a priori be computed.

\acknowledgments
C.~S. and N.~M. thank the F.R.S.-FNRS for financial support.

\appendix

\section{Mean values for completely symmetrized states}
\label{app:meanval}

Let us note $|\phi\rangle$ a $n$-body state completely symmetrized, that is to says that 
\begin{equation}
\label{phiPij}
P_{ij} |\phi\rangle = \pm |\phi\rangle,
\end{equation}
where $P_{ij}$ is the operator permuting particles $i$ and $j$. If $K_i$ is a one-body operator, we can write
\begin{equation}
\label{phiKPij}
\langle \phi | K_i |\phi\rangle = \langle \phi | P^+_{ij} P_{ij} K_i P^+_{ij} P_{ij}|\phi\rangle
= (\pm\langle \phi |) (P_{ij} K_i P^+_{ij}) (\pm|\phi\rangle) = \langle \phi | K_j |\phi\rangle.
\end{equation}
Similarly for a two-body operator $W_{ij}$, we can write
\begin{equation}
\label{phiWPij}
\langle \phi | W_{ij} |\phi\rangle = \langle \phi | W_{kl} |\phi\rangle,
\end{equation}
for any numbers $i\ne j$ and $k\ne l$. So we can write, for instance, that
\begin{align}
\label{K1}
& \langle \phi | \sum_{i=1}^n K_i |\phi\rangle = n \langle \phi | K_1 |\phi\rangle,\\
\label{W12}
& \langle \phi | \sum_{i=1<j}^n W_{ij} |\phi\rangle = \binom{n}{2} \langle \phi | W_{12} |\phi\rangle.
\end{align}

\section{Scaling of $g_A$}
\label{app:ga}

Let us first repeat the analysis of Witten \cite{witten} about mesons, but with quarks in a representation of dimension $d$ of SU($N$). If $J(\bm p)$ is an operator creating a meson with momentum $\bm p$, then the matrix element $\langle 0| J(\bm p)J(-\bm p)| 0\rangle$ is represented by a loop of one quark-line. This line being characterized by $d$ degrees of freedom, the matrix elements is of the order of O$(d)$. On the other hand, we can write
\begin{equation}
\label{JJ}
\langle 0| J(\bm p)J(-\bm p)| 0\rangle = \sum_{k} \frac{a_k^2}{\bm p^2-m_k^2},
\end{equation}
where $m_k$ is the mass of the $k$th meson, and $a_k=\langle 0| J(\bm p)| k\rangle$ is the matrix element for creating the $k$th meson from the vacuum. Since $m_k \sim \textrm{O}(1)$, we must have $a_k \sim \textrm{O}(\sqrt{d})$. In particular this must be true for the pion, and then $f_\pi \propto a_\pi \sim \textrm{O}(\sqrt{d})$. This is in agreement with the results of \cite{sann07}. 

The constant $g_A$ can be computed by the Adler-Weisberger sum rules
\begin{equation}
\label{sumrules}
1-g_A^2=\frac{2 f_\pi^2}{\pi}\int_0^\infty \frac{d\omega}{\omega} \left(\sigma_-(\omega)-\sigma_+(\omega) \right),
\end{equation}
where the pion is assumed to be massless, and $\sigma_\pm (\omega)$ is the total cross section for scattering of $\pi^\pm$ of energy $\omega$ on a up quark at rest. The dominant contribution to this quantity is of the order of $\theta^4$, where $\theta$ is the quark-meson coupling constant \cite{bron93}. The meson-baryon scattering is then a process of the order of $\theta^2\, d$, if the baryon is composed of $d$ quarks. But, if the meson is considered as two quark-lines with one line exchanging a gluon with one quark-line in the baryon, this process is of the order of $2\, \alpha_S\, d$. So, we must have $\theta \sim \sqrt{\alpha_S} \sim \textrm{O}(N^{-1/2})$. Finally, we have
\begin{equation}
\label{approxga}
1-g_A^2 \sim f_\pi^2 \, \theta^4 \sim \textrm{O}(d/N^{2}).
\end{equation}
Within our model, we can identify $d$ with the number $n_q$ of quarks in a baryon (see Sect.~\ref{sec:barwfcol}). So, $1-g_A^2$ is at most of the order of O(1), since $d=N$ for QCD$_\textrm{F}$, $d=N(N-1)/2$ for QCD$_\textrm{AS}$, and $d=N(N+1)/2$ for QCD$_\textrm{Sym}$.


\begin{thebibliography}{99}

\bibitem{hoof} G. 't~Hooft, Nucl.\ Phys.\ B {\bf 72}, 461 (1974). 
\bibitem{manohar} A.~V.~Manohar, hep-ph/9802419, and references therein.
\bibitem{teper} M.~Teper, PoS LATTICE2008, 022 (2008) [arXiv:0812.0085], and references therein.
\bibitem{witten} E. Witten, Nucl. Phys. B \textbf{160}, 57 (1979).
\bibitem{luty93} M. A. Luty and J. March-Russell, Nucl. Phys. B {\bf 426}, 71 (1994); M. A. Luty, Phys. Rev. D {\bf 51}, 2322 (1995).
\bibitem{vene} G. Veneziano, Nucl.\ Phys.\ B {\bf 117}, 519 (1976).
\bibitem{qcdas} A. Armoni, M. Shifman, and G. Veneziano, Nucl. Phys. B \textbf{667}, 170 (2003); 
Phys. Rev. Lett. \textbf{91}, 191601 (2003). 
\bibitem{bolo} S.~Bolognesi, Phys.\ Rev.\ D {\bf 75}, 065030 (2007). 
\bibitem{cohen} T.~D.~Cohen, D.~L.~Shafer, and R.~F.~Lebed, Phys.\ Rev.\ D {\bf 81}, 036006 (2010). 
\bibitem{ande11} J.~R. Andersen \emph{et al.}, Eur. Phys. J. Plus \textbf{126}, 81 (2011).
\bibitem{bolo2} S.~Bolognesi and M.~Shifman, Phys.\ Rev.\ D {\bf 75}, 065020 (2007); R.~Auzzi, S.~Bolognesi, and M.~Shifman, Phys.\ Rev.\ D {\bf 77}, 125029 (2008).
\bibitem{corr} E. Corrigan and P. Ramond, Phys. Lett. B \textbf{87}, 73 (1979). 
\bibitem{rytt06} T. A. Ryttov and F. Sannino, Phys. Rev. D {\bf 73}, 016002 (2006). 
\bibitem{buis10} F. Buisseret and C. Semay, Phys. Rev. D {\bf 82}, 056008 (2010).
\bibitem{afm} B.~Silvestre-Brac, C.~Semay, F.~Buisseret, and F.~Brau, J.\ Math.\ Phys.\ {\bf 51}, 032104 (2010). 
\bibitem{afmdual} B.~Silvestre-Brac and C.~Semay, J. Math. Phys. {\bf 52}, 052107 (2011). 
\bibitem{afmnew} B.~Silvestre-Brac, C.~Semay, and F.~Buisseret, to appear in J. Phys. Math., arXiv:1106.6123.
\bibitem{afmrev} B.~Silvestre-Brac, C.~Semay, and F.~Buisseret, arXiv:1101.5222.
\bibitem{luch91} W. Lucha, F. F. Sch\"oberl, and D. Gromes, Phys. Rep. {\bf 200}, 127 (1991).
\bibitem{gloz96} L. Ya. Glozman and D. O. Riska, Phys. Rep. {\bf 268}, 263 (1996).
\bibitem{pere65} A. M. Perelomov and V. M. Popov, JETP Lett. {\bf 1}, 160 (1965).
\bibitem{stancu} F. Stancu, \emph{Group Theory in Subnuclear Physics} (Clarendon Press, Oxford, 1996).
\bibitem{silv85} B. Silvestre-Brac, J. Physique \textbf{46}, 1087 (1985).
\bibitem{make02} Yu. Makeenko, \textit{Methods of Contemporary Gauge Theory} (Cambridge University Press, Cambridge, 2002).
\bibitem{barn3} C.~Semay, F.~Buisseret, N.~Matagne, and F.~Stancu, Phys. Rev. D \textbf{75}, 096001 (2007);
C.~Semay, F.~Buisseret, and F.~Stancu, Phys. Rev. D \textbf{76}, 116005 (2007). 
\bibitem{luci10} B. Lucini, G. Moraitis, A. Patella, and A. Rago, Phys. Rev. D \textbf{82}, 114510 (2010).
\bibitem{unsa06} M. \"Unsal and L.~G. Yaffe, Phys. Rev. D \textbf{74}, 105019 (2006).
\bibitem{godf85} S. Godfrey and N. Isgur, Phys. Rev. D {\bf 32}, 189 (1985);
S. Capstick and N. Isgur, Phys. Rev. D {\bf 34}, 2809 (1986).
\bibitem{hida11} Y. Hidaka, T. Kojo, L. McLerran, and R. D. Pisarski, Nucl. Phys. A \textbf{852}, 155 (2011). 
\bibitem{gloz98} L. Ya. Glozman, W. Plessas, K. Varga, and R. F. Wagenbrunn, Phys. Rev. D {\bf 58}, 094030 (1998).
\bibitem{bron93} W. Broniowski, M. Lutz and A. Steiner, Phys. Rev. Lett. {\bf 71}, 1787 (1993).
\bibitem{sann07} F. Sannino and J. Schechter, Phys. Rev. D {\bf 76}, 014014 (2007).
\bibitem{wein90} S. Weinberg, Phys. Rev. Lett. {\bf 65}, 1181 (1990).
\bibitem{cher2} A.~Cherman, T.~D.~Cohen, and R.~F.~Lebed, Phys.\ Rev.\ D {\bf 80}, 036002 (2009). 
\bibitem{mata08} N.~Matagne and F.~Stancu, Nucl. Phys. A \textbf{811}, 291 (2008). 

\end{thebibliography}
\end{document}